\documentclass[11pt]{article}%
\usepackage{amsfonts}
\usepackage{amsmath}
\usepackage{amssymb}
\usepackage{graphicx}%
\providecommand{\U}[1]{\protect\rule{.1in}{.1in}}

\begin{document}

\title{On the (Non)Equivalence of the Schr\"{o}dinger and Heisenberg Pictures of
Quantum Mechanics}
\author{Maurice A. de Gosson\\University of Vienna\\Faculty of Mathematics, NuHAG\\Vienna\\maurice.de.gosson@univie.ac.at}
\maketitle

\begin{abstract}
The aim of this short Note is to show that the Schr\"{o}dinger and Heisenberg
pictures of quantum mechanics cannot be equivalent unless one uses a
quantization rule clearly stated by Born and Jordan in their famous 1925
paper. This rule is sufficient and necessary to ensure energy conservation in
Heisenberg's matrix mechanics. It follows, in particular, that Schr\"{o}dinger
and Heisenberg mechanics yield different theories if one quantizes the
Hamiltonian using the Weyl prescription.

\end{abstract}

\section{Introduction}

In the Schr\"{o}dinger picture of quantum mechanics, the operators are
constant (unless they depend explicitly on time), and the states evolve in
time; this evolution is governed by Schr\"{o}dinger's equation%
\begin{equation}
i\hbar\frac{\partial}{\partial t}|\psi\rangle=H|\psi\rangle. \label{sch}%
\end{equation}
In the Heisenberg picture, the state vectors are time-independent operators
that incorporate a dependency on time, while an observable $A_{\mathcal{S}}$
in the Schr\"{o}dinger picture becomes a time-dependent operator
$A_{\mathcal{H}}(t)$ in the Heisenberg picture; this time dependence satisfies
the Heisenberg equation%
\begin{equation}
i\hbar\frac{dA_{\mathcal{H}}}{dt}=i\hbar\frac{\partial A_{\mathcal{H}}%
}{\partial t}+[A_{\mathcal{H}},H_{\mathcal{H}}]. \label{hei}%
\end{equation}

Both theories are related as follows \cite{Messiah,Schiff}: Let $(U(t,t_{0}))$
be the propagator determined by Schr\"{o}dinger's equation:%
\begin{equation}
i\hbar\frac{d}{dt}U(t,t_{0})=\widehat{H}U(t,t_{0})\text{ \ , \ }U(t_{0}%
,t_{0})=I_{\mathrm{d}}; \label{ut}%
\end{equation}
the operators $U(t,t_{0})$ are unitary on $L^{2}(\mathbb{R})$; a ket
\begin{equation}
|\psi_{\mathcal{S}}(t)\rangle=U(t,t_{0})|\psi_{\mathcal{S}}(t_{0}%
)\rangle\label{psis}%
\end{equation}
in the Schr\"{o}dinger picture becomes, in the Heisenberg picture, the
constant ket
\begin{equation}
|\psi_{\mathcal{H}}\rangle=U(t,t_{0})^{\ast}|\psi_{\mathcal{S}}(t)\rangle
=|\psi_{\mathcal{S}}(t_{0})\rangle\label{psih}%
\end{equation}
whereas an observable $A_{\mathcal{S}}$ becomes
\begin{equation}
A_{\mathcal{H}}(t)=U(t,t_{0})^{\ast}A_{\mathcal{S}}U(t,t_{0}); \label{ah}%
\end{equation}
in particular the Hamiltonian is%
\begin{equation}
H_{\mathcal{H}}(t)=U(t,t_{0})^{\ast}H_{\mathcal{S}}U(t,t_{0}). \label{hh}%
\end{equation}

We are going to see that this relation implies that $H_{\mathcal{H}}$ and
$H_{\mathcal{S}}$ are identical and must thus be quantized using the
\emph{same} rules. The proof of this essential property is based on Born and
Jordan's analysis of Heisenberg's foundational paper \cite{hei}. The idea is
that the Heisenberg picture requires that the Hamiltonian operator
$H_{\mathcal{H}}(t)$ must be a constant of the motion, which implies, taking
(\ref{hh}) into account, that $H_{\mathcal{H}}(t)=H_{\mathcal{S}}$ for all
times $t$. A consequence of this property is that if we believe that
Heisenberg's \textquotedblleft matrix mechanics\textquotedblright\ is correct
and is equivalent\footnote{See however \cite{madrid,muller} for a discussion
of the \textquotedblleft myth of equivalence\textquotedblright.} to
Schr\"{o}dinger's theory , then the Hamiltonian operator appearing in the
Schr\"{o}dinger equation (\ref{sch}) \emph{must} be quantized using the
Born--Jordan rules, which are different from (for instance) the Weyl
quantization rule.

We begin by shortly exposing the main arguments in Born and Jordan's paper
\cite{bj}.

\section{The Born--Jordan paper in a nutshell}

We are working in a one-dimensional configuration space; the discussion is
generalized \textit{mutatis mutandis} in \cite{bjh}. Following Heisenberg's
paper \cite{hei} Born and Jordan considered in \cite{bj}\footnote{English
translation in \cite{sources}.} square infinite matrices%
\begin{equation}
\mathbf{a}=(a(n,m))=%
\begin{pmatrix}
a(00) & a(01) & a(02) & \cdot\cdot\cdot\\
a(10) & a(11) & a(12) & \cdot\cdot\cdot\\
a(20) & a(21) & a(22) & \cdot\cdot\cdot\\
\cdot\cdot\cdot & \cdot\cdot\cdot & \cdot\cdot\cdot & \cdot\cdot\cdot
\end{pmatrix}
\label{1}%
\end{equation}
where the $a(nm)$ are what they call \textquotedblleft ordinary
quantities\textquotedblright, \textit{i.e.} scalars; we will call these
infinite matrices (for which we always use boldface letters)
\emph{observables}. In particular Born and Jordan introduce momentum and
position observables $\mathbf{p}$ and $\mathbf{q}$ and matrix functions
$\mathbf{H}(\mathbf{p,q)}$ of these observables, which they call
\textquotedblleft Hamiltonians\textquotedblright. Following Heisenberg, they
assume that the equations of motion for $\mathbf{p}$ and $\mathbf{q}$ are
formally the same as in classical theory, namely%
\begin{equation}
\mathbf{\dot{q}}=\frac{\partial\mathbf{H}}{\partial\mathbf{p}}\text{ \ ,
\ }\mathbf{\dot{p}}=-\frac{\partial\mathbf{H}}{\partial\mathbf{q}};\label{3}%
\end{equation}
limiting themselves deliberately to Hamiltonians which are polynomials in the
observables $\mathbf{p}$, $\mathbf{q}$, that is linear combinations of
monomials which are products of terms
\begin{equation}
\mathbf{H=p}^{s}\mathbf{q}^{r}\label{4}%
\end{equation}
they define the derivatives in (\ref{3}) by the formulas, and show that the
observables $\mathbf{p}$ and $\mathbf{q}$ satisfy the commutation relation%
\begin{equation}
\mathbf{pq}-\mathbf{qp}=\frac{h}{2\pi i}\mathbf{1}\label{2}%
\end{equation}
where $\mathbf{1}$ is the identity matrix; more generally,%
\begin{equation}
\mathbf{p}^{m}\mathbf{q}^{n}-\mathbf{q}^{n}\mathbf{p}^{m}=m\frac{h}{2\pi
i}\sum_{\ell=0}^{n-1}\mathbf{q}^{n-1-\ell}\mathbf{p}^{m-1}\mathbf{q}^{\ell
}.\label{7}%
\end{equation}
Born and Jordan next proceed to derive the fundamental laws of quantum
mechanics. In particular, pursuing their analogy with classical mechanics,
they want to prove that energy is conserved; identifying the values of the
Hamiltonian $\mathbf{H}$ with the energy of the system, they impose the
condition $\mathbf{\dot{H}=0}$ and show that this condition requires that%
\begin{equation}
\mathbf{\dot{q}}=\frac{2\pi i}{h}(\mathbf{Hq-qH)}\text{ \ , \ }\mathbf{\dot
{p}}=\frac{2\pi i}{h}(\mathbf{Hp-pH).}\label{10}%
\end{equation}
Comparing with the Hamilton-like equations (\ref{3})\ this condition is in
turn equivalent to
\begin{equation}
\mathbf{Hq-qH}=\frac{h}{2\pi i}\frac{\partial\mathbf{H}}{\partial\mathbf{p}%
}=\text{ \ , }\mathbf{Hp-pH}=-\frac{h}{2\pi i}\frac{\partial\mathbf{H}%
}{\partial\mathbf{q}}.\label{11}%
\end{equation}
Now comes the crucial step. Given a classical Hamiltonian $H(p,q)=p^{s}q^{r}$
they ask how one should choose the observable $\mathbf{H}(\mathbf{p}%
,\mathbf{q})$. Using the commutation formula (\ref{7}) Born and Jordan show
that the only possible choice satisfying the conditions (\ref{11}) is%
\begin{equation}
\mathbf{H}(\mathbf{p},\mathbf{q})=\frac{1}{s+1}\sum_{\ell=0}^{s}%
\mathbf{p}^{s-\ell}\mathbf{q}^{r}\mathbf{p}^{\ell}\label{12}%
\end{equation}

\section{Quantum mechanics and Born--Jordan quantization}

Born and Jordan thus proved --rigorously-- that the only way to quantize
polynomials in a way consistent with Heisenberg's ideas was to use the rule%
\begin{equation}
p^{s}q^{r}\overset{\mathrm{BJ}}{\longrightarrow}\frac{1}{s+1}\sum_{\ell=0}%
^{s}\mathbf{p}^{s-\ell}\mathbf{q}^{r}\mathbf{p}^{\ell}=\frac{1}{r+1}\sum
_{j=0}^{r}\mathbf{q}^{r-j}\mathbf{p}^{s}\mathbf{q}^{j} \label{bj1}%
\end{equation}
(the equality following from the commutation relation (\ref{7}). In their
subsequent publication \cite{bjh}\footnote{English translation in
\cite{sources}.} with Heisenberg they show that their constructions extend
\textit{mutatis mutandis} to systems with an arbitrary number of degrees of
freedom. We will call this rule (and its extension to higher dimensions) the
\emph{Born--Jordan quantization rule}. Weyl \cite{Weyl} proposed,
independently, some time later (1926) another rule leading, for monomials, to
the replacement of (\ref{bj1}) with
\begin{equation}
p^{s}q^{r}\overset{\mathrm{Weyl}}{\longrightarrow}\frac{1}{2^{s}}\sum_{\ell
=0}^{s}%
\begin{pmatrix}
s\\
\ell
\end{pmatrix}
\mathbf{p}^{s-\ell}\mathbf{q}^{r}\mathbf{p}^{\ell}. \label{w2}%
\end{equation}
It turns out that both rules coincide when $m+n\leq2$, but they do not for
higher values of $m+n$. For instance%
\[
p^{2}q\overset{\mathrm{BJ}}{\longrightarrow}\frac{1}{3}(\mathbf{p}%
^{2}\mathbf{q}+\mathbf{pqp}+\mathbf{qp}^{2})\text{ \ , \ }p^{2}%
q\overset{\mathrm{Weyl}}{\longrightarrow}\frac{1}{4}(\mathbf{p}^{2}%
\mathbf{q}+2\mathbf{pqp}+\mathbf{qp}^{2}).
\]
Both quantizations are thus not equivalent; as Kauffmann \cite{Kauffmann}
observes, Weyl's rule is the single most symmetrical operator ordering,
whereas the Born--Jordan quantization is the equally weighted average of all
the operator orderings.

These facts have the following consequence: if we insist that the Heisenberg
and Schr\"{o}dinger pictures be \emph{equivalent}, then we \emph{must}
quantize the Hamiltonian in Schr\"{o}dinger's equation using Born--Jordan
quantization. In fact, recall from formula (\ref{hh}) that the Heisenberg and
Schr\"{o}dinger Hamiltonians $H_{\mathcal{H}}$ and $H_{\mathcal{S}}$ are
related by
\[
H_{\mathcal{H}}(t)=U(t,t_{0})^{\ast}H_{\mathcal{S}}U(t,t_{0}).
\]
Since $H_{\mathcal{H}}(t)$ is a constant of the motion we have $H_{\mathcal{H}%
}(t)=H_{\mathcal{H}}(t_{0})$ and hence $H_{\mathcal{H}}(t)=H_{\mathcal{S}}$ so
the Heisenberg and Schr\"{o}dinger Hamiltonians $H_{\mathcal{H}}(t)$ and
$H_{\mathcal{H}}$ must be identical. But the condition $H_{\mathcal{H}%
}(t)=H_{\mathcal{H}}(t_{0})=H_{\mathcal{H}}$ means that $H_{\mathcal{H}}$ and
hence $H_{\mathcal{S}}$ must be quantized using the Born--Jordan prescription.

An obvious consequence of these considerations is that if one uses in the
Schr\"{o}dinger picture the Weyl quantization rule (or any other quantization
rule), we obtain two different renderings of quantum mechanics. This
observation seems to be confirmed by Kauffmann's \cite{Kauffmann} interesting
discussion of the non-physicality of Weyl quantization.

We have been considering the quantization of polynomials for simplicity; in de
Gosson and Luef \cite{GoLuBJ} and de Gosson \cite{TRANSAM} we have shown in
detail how to Born--Jordan quantize arbitrary functions of the position and
momentum variables.


\begin{thebibliography}{99}                                                                                               %
\bibitem {bj}M. Born, P. Jordan, Zur Quantenmechanik, \textit{Z. Physik}
\textbf{34}, 858--888 (1925)

\bibitem {bjh}M. Born, W. Heisenberg, and P. Jordan, Zur Quantenmechanik II,
\textit{Z. Physik} \textbf{35}, 557-615 (1925)

\bibitem {GoLuBJ}M. de Gosson, F. Luef, Preferred quantization rules:
Born-Jordan versus Weyl. The pseudo-differential point of view. \textit{J.
Pseudo-Differ. Oper. Appl.} \textbf{2} (2011), no. 1, 115--139

\bibitem {TRANSAM}M. de Gosson, Symplectic covariance properties for Shubin
and Born-Jordan pseudo-differential operators. \textit{Trans. Amer. Math.
Soc.} \textbf{365} (2013), no. 6, 3287--3307

\bibitem {hei}W. Heisenberg, \"{U}ber quantentheoretische Umdeutung
kinematischer und mechanischer Beziehungen, \textit{Z. Physik} \textbf{33},
879--893 (1925)

\bibitem {Kauffmann}S. K. Kauffmann, Unambiguous Quantization from the Maximum
Classical Correspondence that Is Self-consistent: The Slightly Stronger
Canonical Commutation Rule Dirac Missed, \textit{Found. Phys}. \textbf{41},
no. 5, 805--819 (2011)

\bibitem {madrid}C. M. Madrid Casado, A brief history of the mathematical
equivalence between the two quantum mechanics, Latin American Journal of
Physics Education \textbf{2}(2), 104--108 (2008)

\bibitem {muller}F. A. Muller, The equivalence myth of quantum mechanics; Part
I: \textit{Stud. Hist. Phil. Mod. Phys}. \textbf{28}(1), 35--61 (1997); Part
II: \textit{Stud. Hist. Phil. Mod. Phys}. \textbf{28}(2), 219--241 (1997)

\bibitem {Messiah}A. Messiah, Quantum Mechanics (Vol. I), English translation
from French by G. M. Temmer. North Holland, John Wiley \& Sons, 1966

\bibitem {Schiff}L. I. Schiff, Quantum Mechanics. McGraw-Hill, New York, 3d
edition, 1968

\bibitem {sources}B. L. van der Waerden, ed. Sources of quantum mechanics.
Courier Dover Publications, 2007

\bibitem {Weyl}H. Weyl, Quantenmechanik und Gruppentheorie, \textit{Z. Physik}
\textbf{46} (1927)
\end{thebibliography}
\end{document}